\documentstyle[twoside,fleqn,npb,epsfig]{article}
\newcommand{\AmS}{{\protect\the\textfont2
  A\kern-.1667em\lower.5ex\hbox{M}\kern-.125emS}}
\newcommand{\be}{\begin{equation}}
\newcommand{\ee}{\end{equation}}
\newcommand{\bea}{\begin{eqnarray}}
\newcommand{\eea}{\end{eqnarray}}
\newcommand{\bean}{\begin{eqnarray*}}
\newcommand{\eean}{\end{eqnarray*}}
\newcommand{\ba}{\begin{array}}
\newcommand{\ea}{\end{array}}

\newcommand{\slashl}[1]{\not{\!\!#1}}
\newcommand{\slashs}[1]{\not{\!#1}}

\title{$B$ meson light-cone wavefunctions in the heavy quark limit\thanks{Talk presented by K. Tanaka
at the International Symposium Radcor 2002 and Loops and Legs 2002, Kloster Banz, September 8-13, 2002.}}

\author{H. Kawamura\address{Deutsches Elektronen-Synchrotron, DESY\\
Platanenallee 6, D 15738 Zeuthen, Germany},
        J. Kodaira\address{Department of Physics, Hiroshima University\\
Higashi-Hiroshima 739-8526, Japan}%
\thanks{Supported in part by the Monbu-kagaku-sho Grant-in-Aid
for Scientific Research No.C-13640289.},
        C.-F. Qiao$^{\rm b}$%
\thanks{Supported by the Grant-in-Aid of JSPS committee.},
        K. Tanaka\address{Department of Physics, Juntendo University\\
Inba-gun, Chiba 270-1695, Japan}}

\begin{document}

\begin{abstract} 
We present a systematic study of the $B$ meson light-cone wavefunctions
in QCD in the heavy-quark limit.  
We construct model-independent formulae 
for the light-cone wavefunctions 
in terms of independent 
dynamical degrees of freedom,
which exactly satisfy the QCD equations of motion 
and constraints from heavy-quark symmetry.
The results demonstrate novel behaviors of longitudinal as well
as transverse momentum distribution in the $B$ mesons.
\end{abstract}

\maketitle


Recently systematic methods based on the QCD factorization
have been developed for the exclusive $B$ meson decays
into light mesons \cite{Beneke:2000ry}
(see also Ref. \cite{kls}).
Essential ingredients in this approach are
the light-cone distribution amplitudes for the participating mesons,
which express nonperturbative long-distance contribution
to the factorized amplitudes.
The light-cone distribution amplitudes
describe the probability
amplitudes 
to find the meson in a state
with the constituents carrying definite 
light-cone momentum
fraction,
and thus are process-independent quantity.
{}For the light mesons ($\pi$, $K$, 
$\rho$, 
$K^{*}$, etc.)
appearing in the final state,
systematic model-independent study of the light-cone distributions 
exists for both leading 
and higher twists \cite{Braun:1990iv}.
On the other hand, 
the light-cone distribution amplitudes for
the $B$ mesons 
are 
not well-known 
at present 
and they 
provide a major source of uncertainty in the calculations of 
the decay rates.

By definition, 
the 
distribution amplitudes are obtained from 
light-cone 
wavefunctions at 
(almost) 
zero 
transverse separation 
of the constituents,
$\phi(x) \sim \int_{k_{T}^{2} < \mu^{2}} d^{2}k_{T} \Phi(x, \mbox{\boldmath $k$}_{T})$.
The light-cone wavefunctions with transverse momentum dependence
are also necessary for computing the power corrections to the exclusive
amplitudes,
and 
for estimating the transition form factors for $B \rightarrow D$, $B
\rightarrow \pi$, etc,
which constitute another type of long-distance contributions 
appearing 
in
the factorization
approaches for the exclusive $B$ meson decays.

In this work \cite{KKQT,KKQT2}, 
we demonstrate that, in the heavy-quark limit
relevant for the factorization approaches
for the exclusive $B$ meson decays,
the $B$ meson light-cone wavefunctions obey
exact differential equations, which are based on
heavy-quark symmetry and 
the QCD equations of motion.
As solution of those differential equations,
we derive the model-independent formulae
for the light-cone wavefunctions,
which 
involve not only the leading Fock-states with a minimal number of (valence) partons
but also the higher Fock-states
with additional dynamical gluons.
%

The light-cone wavefunctions are related to the usual Bethe-Salpeter
wavefunctions at equal
light-cone time $z^{+}= (z^{0}+z^{3})/\sqrt{2}$.
In the heavy-quark limit,
the quark-antiquark light-cone wavefunctions $\tilde{\Phi}_{\pm}(t,z^2)$ of
the
$B$ mesons 
can be introduced 
in terms of vacuum-to-meson matrix element of
nonlocal operators in the 
the heavy-quark effective theory (HQET)
\cite{sachrajda,KKQT2}:
\bea
\lefteqn{\langle 0 | \bar{q}(z) \Gamma h_{v}(0) |\bar{B}(p) \rangle
 = - \frac{i f_{B} M}{2} {\rm Tr}
 \left[ \gamma_{5}\Gamma \frac{1 + \slashs{v}}{2}\right.}\nonumber\\
&&\!\!\!\times \left. \left\{ \tilde{\Phi}_{+}(t,z^2) - \slashs{z} \frac{\tilde{\Phi}_{+}(t,z^2)
 -\tilde{\Phi}_{-}(t,z^2)}{2t}\right\} \right].
 \label{phi}
\eea
Here 
$z^{\mu}=(0, z^{-}, \mbox{\boldmath $z$}_{T})$, $z^{2}= - \mbox{\boldmath $z$}_{T}^{2}$,
$v^{2} = 1$, $t=v\cdot z$, and $p^{\mu} = Mv^{\mu}$
is the 4-momentum of the $B$ meson with mass $M$.
$h_{v}(x)$ denotes the effective $b$-quark field,
$b(x) \approx \exp(-im_{b} v\cdot x)h_{v}(x)$,
and is subject to the on-shell constraint,
$\slashl{v} h_{v} = h_{v}$ \cite{Neubert:1994mb}.
$\Gamma$ is a generic Dirac matrix and,
here and in the following, the path-ordered gauge
factors are implied in between the constituent fields.
$f_{B}$ is the decay constant defined as usually as
%
$\langle 0 | \bar{q}(0) \gamma^{\mu}\gamma_{5} h_{v}(0) |\bar{B}(p) \rangle
   = i f_{B} M v^{\mu}$,
%
so that $\tilde{\Phi}_{\pm}(t=0,z^2=0) = 1$.
Eq. (\ref{phi}) is the most general parameterization compatible with Lorentz
invariance and the heavy-quark limit.

Higher Fock components in the $B$ mesons are described by
multi-particle wavefunctions. 
We explicitly deal with quark-antiquark-gluon three-particle 
wavefunctions, defined as \cite{KKQT}
\bea
 \lefteqn{\langle 0 | \bar{q} (z) \, g G_{\mu\nu} (uz)\, z^{\nu}
      \, \Gamma \, h_{v} (0) | \bar{B}(p) \rangle}\nonumber \\
 & &= \frac{1}{2}\, f_B M \, {\rm Tr}\, \left[ \, \gamma_5\,
      \Gamma \,
        \frac{1 + \slashs{v}}{2}\,  
\biggl\{ ( v_{\mu}\slashs{z}
         - t \, \gamma_{\mu} )\  \right. \nonumber\\
&&\times\left( \tilde{\Psi}_A (t,u)
   - \tilde{\Psi}_V (t,u) \right)
- i \, \sigma_{\mu\nu} z^{\nu}\,
           \tilde{\Psi}_V (t,u)
\nonumber\\
      & &- \left. z_{\mu} \, \tilde{X}_A (t,u)\,
+ \frac{z_{\mu}}{t} \, \slashs{z} \,\tilde{Y}_A \,(t\,,\,u)
\biggr\} \, \right] + \ldots ,
\label{3elements0}
\eea
where the ellipses stand for the terms which involve
one or more powers of $z^2$ and are irrelevant for the
present work.
We have the four functions
$\tilde{\Psi}_V , \tilde{\Psi}_A ,
\tilde{X}_A$ and $\tilde{Y}_A$
as the independent three-particle wavefunctions in the heavy-quark limit.

The QCD equations of motion
impose a set of relations between the above wavefunctions \cite{KKQT,KKQT2}. 
They can be derived most directly 
from the exact identities between the nonlocal operators:
\bea
 \lefteqn{\frac{\partial}{\partial x^{\mu}}
 \bar{q}(x) \gamma^{\mu} \Gamma h_{v}(0)}\nonumber\\
  &&=
i \int_{0}^{1}duu \ \bar{q}(x) gG_{\mu \nu}(ux) x^{\nu}
  \gamma^{\mu}\Gamma h_{v}(0) \ ,
 \label{id1} \\
\lefteqn{ v^{\mu}\frac{\partial}{\partial x^{\mu}}
 \bar{q}(x) \Gamma h_{v}(0)}\nonumber\\
 &&= 
i \int_{0}^{1}du (u-1)\ 
\bar{q}(x) gG_{\mu \nu}(ux) 
 v^{\mu}x^{\nu}\Gamma h_{v}(0) 
\nonumber \\
 &&+
 v^{\mu}\left.
   \frac{\partial}{\partial y^{\mu}}\
  \bar{q}(x+y) \Gamma h_{v}(y)\right|_{y \rightarrow 0}\ ,
\label{id2}
\eea
%
where $G_{\mu \nu}$ is the gluon field strength tensor, and
we have used the equations of motion
$\slashl{D}q=0$ 
and $v \cdot D h_{v} = 0$
with $D_{\mu}= \partial_{\mu} - igA_{\mu}$ the covariant derivative.
%
%
Taking the 
matrix element with $x_{\mu} \rightarrow z_{\mu}$,
the LHS of these identities
yield
$\tilde{\Phi}_{+}(t, z^2 )$, $\tilde{\Phi}_{-}(t, z^2 )$ defined in Eq. (\ref{phi}) 
and their derivatives,
$\partial \tilde{\Phi}_{\pm}(t, z^{2})/\partial t$ 
and 
$\partial \tilde{\Phi}_{\pm}(t, z^{2})/\partial z^{2}$.
%
The terms in the RHS, which are given by integral of quark-antiquark-gluon operator,
are expressed by the three-particle wavefunctions of Eq. (\ref{3elements0}).
The last term of Eq. (\ref{id2}), the derivative over the total translation,
yields 
$\tilde{\Phi}_{\pm}(t, z^2 )$ multiplied by
\be 
  \bar{\Lambda} = M - m_{b} =
 \frac{iv\cdot \partial \langle 0| \bar{q} \Gamma h_{v} |\bar{B}(p) \rangle}
  {\langle 0| \bar{q} \Gamma h_{v} |\bar{B}(p) \rangle}\ .
\label{lambda}
\ee
This is the usual ``effective mass'' of meson states in the 
HQET \cite{Neubert:1994mb}.

Substituting all the 
Dirac matrices for $\Gamma$,
we obtain the four independent constraint equations between the two- and three-particle wavefunctions
from Eqs. (\ref{id1}) and (\ref{id2}).
We solve this system of equations for the 
relevant 
two cases:
(i) In the light-cone limit $z^2 \rightarrow 0$ ($\mbox{\boldmath $z$}_{T} \rightarrow 0$), 
with fully taking into account
the contribution due to the three-particle wavefunctions.
The solution gives exact model-independent representaions
for the light-cone distribution amplitudes in terms of independent dynamical degrees
of freedom.
(ii) For 
$z^2 \neq 0$ but in the approximation neglecting the
contribution of the three-particle wavefunctions. 
The solution gives exact analytic formulae
for the light-cone wavefunctions with transverse momentum dependence within the valence Fock-states.

Now we discuss the case (i) in detail \cite{KKQT}. In the light-cone limit,
the light-cone wavefunctions of Eq. (\ref{phi})
reduce to 
the light-cone distribution amplitudes $\tilde{\phi}_{\pm}(t)$
as \cite{Grozin:1997pq}
\begin{equation}
\tilde{\phi}_{\pm}(t)=\tilde{\Phi}_{\pm}(t, z^{2}=0)\ ,
\label{sphi}
\end{equation}
and we also introduce the
shorthand notations, 
$\tilde{\phi}_{\pm}'(t) \equiv d \tilde{\phi}_{\pm}(t)/dt$ and 
$\partial \tilde{\phi}_{\pm}(t)/\partial z^{2}\equiv
\partial \tilde{\Phi}_{\pm}(t, z^{2})/\partial z^{2}
|_{z^{2} \rightarrow 0}$,
%
%
which denote the derivatives with respect to the longitudinal and transverse separations,
respectively.
The first identity (\ref{id1}) yields the two 
equations:
%
\bea
 \lefteqn{\tilde{\phi}_{-}'(t)- \frac{1}{t}\left(\tilde{\phi}_{+}(t)
 - \tilde{\phi}_{-}(t)\right)}\nonumber\\
 &&=
 2 t \int_0^1 du\, u\, ( \tilde{\Psi}_A (t,u) - \tilde{\Psi}_V (t,u)) \ ,
 \label{de1} \\
 \lefteqn{\tilde{\phi}_{+}'(t)
 -\tilde{\phi}_{-}'(t) - \frac{1}{t}\left(\tilde{\phi_{+}}(t)
 - \tilde{\phi}_{-}(t)\right) + 4t
 \frac{\partial \tilde{\phi}_{+}(t)}{\partial z^{2}}} \nonumber \\
=&&\!\!\!\!\!\!\!\!\!\!\! 2  t \!\int_0^1 \!\!\! du\, u\, ( \tilde{\Psi}_A (t,u)
  + 2\, \tilde{\Psi}_V (t,u) + \tilde{X}_A (t,u)),
\label{de2}
\eea
and similarly the second identity (\ref{id2}) yields
%
\bea
\lefteqn{\tilde{\phi}_{+}'(t)- \frac{1}{2t}\left(\tilde{\phi}_{+}(t)
 - \tilde{\phi}_{-}(t)\right)  + i \bar{\Lambda}\tilde{\phi}_{+}(t)
 + 2t \frac{\partial \tilde{\phi}_{+}(t)}{\partial z^{2}}} \nonumber \\
  &&=  t\, \int_0^1 du\,(u-1)\,
           ( \tilde{\Psi}_A (t,u) + \tilde{X}_A (t,u) )\ , \label{de3} \\
 \lefteqn{\tilde{\phi}_{+}'(t)
  -\tilde{\phi}_{-}'(t) +\left(i \bar{\Lambda} -\frac{1}{t}\right)
 \left(\tilde{\phi}_{+}(t) - \tilde{\phi}_{-}(t)\right)}\nonumber\\
&& + 2t \left(
  \frac{\partial \tilde{\phi}_{+}(t)}{\partial z^{2}}-
  \frac{\partial \tilde{\phi}_{-}(t)}{\partial z^{2}}\right) \nonumber \\
  &&=  2t \,\int_0^1 du\,(u-1)\,
           ( \tilde{\Psi}_A (t,u) + \tilde{Y}_A (t,u) ).
\label{de4}
\eea
These Eqs. (\ref{de1})-(\ref{de4}) are exact in QCD
in the heavy-quark limit.

Important observation is that we can eliminate the term
$\partial \tilde{\phi}_{+} (t)/\partial z^2$
by combining Eqs. (\ref{de2}) and (\ref{de3}).
%
%
The resulting equation, combined with Eq. (\ref{de1}), gives 
a system of two differential equations which involve the 
degrees of freedom along the light-cone only.
By going over to the momentum space
by $\tilde{\phi}_{\pm}(t) = \int d\omega \ e^{-i \omega t}
  \phi_{\pm}(\omega)$, and $\tilde{F}(t, u) = \int d\omega d \xi \
  e^{-i(\omega  + \xi u)t} F(\omega, \xi)$
with $F=\{ \Psi_{V}, \Psi_{A}, X_{A}, Y_{A} \}$,
the corresponding differential equations read \cite{KKQT}
\bea
 \omega \frac{d \phi_{-}(\omega)}{d \omega}
  &+& \phi_{+}(\omega) = I(\omega)\ ,
  \label{mde1} \\
  \left(\omega - 2 \bar{\Lambda}\right)\phi_{+}(\omega)
 &+& \omega \phi_{-}(\omega) = J(\omega) \ , \label{mde2}
\eea
where 
$I(\omega)$ and $J(\omega)$ 
denote the 
``source'' terms
due to three-particle wavefunctions as
%
\bea
\lefteqn{I(\omega)
 \! = 2\frac{d}{d\omega}
   \int_{0}^{\omega} \!\!\!\! d\rho \!\! \int_{\omega - \rho}^{\infty} \!\!\! \frac{d\xi}{\xi}
\frac{\partial}{\partial \xi}\left[ \Psi_{A}(\rho, \xi)
   - \Psi_{V}(\rho, \xi)\right],}\nonumber\\
\lefteqn{J(\omega) \!= \! -2\frac{d}{d\omega}
  \int_{0}^{\omega} \!\!\!\! d\rho \!\! \int_{\omega - \rho}^{\infty} \!\!\! \frac{d\xi}{\xi}
\left[ \Psi_{A}(\rho, \xi) + X_{A}(\rho, \xi)\right]}
  \nonumber \\
  && -4 \int_{0}^{\omega}d\rho \int_{\omega - \rho}^{\infty}\frac{d\xi}{\xi}
  \frac{\partial \Psi_{V}(\rho, \xi)}{\partial \xi} \ . \label{sj}
\eea
%
Eqs. (\ref{mde1}), (\ref{mde2})
can be solved for $\phi_{+}(\omega)$ and $\phi_{-}(\omega)$.
The boundary conditions are specified as $\phi_{\pm}(\omega) = 0$
for $\omega < 0$ or $\omega \rightarrow \infty$, 
because $\omega v^{+}$ has the meaning of the light-cone projection
$k^{+}$ of the light-antiquark momentum in the $B$ meson, and 
the normalization condition is 
$\int_{0}^{\infty}d\omega \phi_{\pm}(\omega)= 
\tilde{\Phi}_{\pm}(0,0)
= 1$.
Obviously, the solution can be decomposed into two pieces as
\be
  \phi_{\pm}(\omega) = \phi_{\pm}^{(W)}(\omega) 
  + \phi_{\pm}^{(g)}(\omega) \ ,
\label{decomp}
\ee
where $\phi_{\pm}^{(W)}(\omega)$ are the solution 
with 
$I(\omega)=J(\omega)=0$,
which corresponds to the ``Wandzura-Wilczek approximation \cite{Braun:1990iv}'' $\Psi_{V}=\Psi_{A}=X_{A}=Y_{A}=0$.
$\phi_{\pm}^{(g)}(\omega)$ denote the pieces induced by the source terms $I(\omega)$ and $J(\omega)$.

We are able to obtain 
the analytic solution for the Wandzura-Wilczek part 
as 
%
\be
 \phi_{\pm}^{(W)}(\omega) = \frac{\bar{\Lambda} \pm (\omega-\bar{\Lambda})}{2 \bar{\Lambda}^{2}} 
 \theta(\omega)\theta(2 \bar{\Lambda} - \omega)\ .
\label{solm}
\ee
%
Moreover, the solution for $\phi_{\pm}^{(g)}$ can be obtained
straightforwardly, and reads ($\omega \ge 0$):
\bea
 \phi_{+}^{(g)}(\omega) &=& \frac{\omega}{2\bar{\Lambda}}
{\cal G}
(\omega)\ ,
  \label{solpg} \\
  \phi_{-}^{(g)}(\omega) &=&
  \frac{2\bar{\Lambda}-\omega}{2\bar{\Lambda}}
{\cal G}
(\omega)
 + \frac{J(\omega)}{\omega}\ ,\label{solmg}\\
%
%
{\cal G}(\omega) &=& \theta(2\bar{\Lambda}-\omega)
 \left\{\int_{0}^{\omega}d\rho \frac{K(\rho)}{2\bar{\Lambda} - \rho}
 -\frac{J(0)}{2\bar{\Lambda}}\right\} \nonumber\\
&& - \theta(\omega - 2\bar{\Lambda}) \int_{\omega}^{\infty}
  d\rho \frac{K(\rho)}{2\bar{\Lambda} - \rho} \nonumber \\
   &&- \int_{\omega}^{\infty}d\rho
  \left( \frac{K(\rho)}{\rho} + \frac{J(\rho)}{\rho^{2}} \right) \ ,
\label{Phi}
\eea
with
%
$K(\rho) = I(\rho) + \left[ 1/(2\bar{\Lambda}) -
    d/d\rho\right]J(\rho)$.
%
The solution (\ref{decomp}) with Eqs. (\ref{solm})-(\ref{Phi})
is exact, and
reveals that 
$\phi_{\pm}$ 
contain the three-particle contributions.
This is also visualized explicitly in terms of 
the Mellin moments
$\langle \omega^{n} \rangle_{\pm} \equiv \int d\omega \ \omega^{n} \phi_{\pm}(\omega)$ 
($n= 0, 1, 2, \cdots$).
Here we present some examples for a few low moments:
$\langle \omega \rangle_{\pm} = (3\pm 1)\bar{\Lambda}/3$, and 
\be
\langle \omega^{2} \rangle_{\pm} = \frac{4\pm 2}{3} \bar{\Lambda}^{2}+\frac{1}{3}\lambda_{E}^{2}
 +\frac{1}{3}\lambda_{H}^{2}  \pm \frac{1}{3}\lambda_{E}^{2}\ ,
\label{mome2}
\ee
where $\lambda_{E}$ and $\lambda_{H}$ 
are due to $\phi_{\pm}^{(g)}$, and 
are related to the chromoelectric and chromomagnetic fields
in the $B$ meson rest frame as
%
$\langle 0 |\bar{q} g \mbox{\boldmath $E$}\cdot\mbox{\boldmath $\alpha$}
 \gamma_{5}h_{v} |\bar{B}(\mbox{\boldmath $p$}=0)\rangle
 = f_{B}M \lambda_{E}^{2}$,
$\langle 0 |\bar{q} g \mbox{\boldmath $H$}\cdot\mbox{\boldmath $\sigma$}
 \gamma_{5}h_{v} |\bar{B}(\mbox{\boldmath $p$}=0)\rangle
 = if_{B}M \lambda_{H}^{2}$.
%
Our solution (\ref{decomp}) allows us to further derive the analytic formulae 
for the general moment $n$ in terms of matrix element of local
two- and three-particle operators with dimension $n+3$ (see Ref. \cite{KKQT} for the detail).


The behavior of 
Eq. (\ref{phi})
for a fast-moving meson, $t = v\cdot z \rightarrow \infty$, 
shows that 
$\phi_{+}$ is of
leading-twist 
whereas $\phi_{-}$ has subleading twist;
the three-particle contributions to 
the leading-twist 
$\phi_{+}$ 
are in contrast with the case of the light mesons \cite{Braun:1990iv},
where the leading-twist amplitudes correspond to
the valence Fock component,
while the higher-twist amplitudes involve 
multi-particle states.
We note that
there exists
an 
estimate
$\lambda_{E}^{2}/\bar{\Lambda}^{2} = 0.36 \pm 0.20$,
$\lambda_{H}^{2}/\bar{\Lambda}^{2} = 0.60 \pm 0.23$
by QCD sum rules \cite{Grozin:1997pq} (see Eq. (\ref{mome2})).
This might suggest that, in the $B$ mesons,
the three-particle contributions
could play important roles even in the leading twist level.
In this connection, 
the shape of our Wandzura-Wilczek
contributions (\ref{solm}),
which are determined uniquely in analytic form in terms of $\bar{\Lambda}$, 
is rather different from various ``model'' distribution amplitudes that have been used
in the existing literature: One exapmle of such models is 
$\phi^{GN}_{+}(\omega) = (\omega/\omega_{0}^{2})
e^{- \omega/\omega_{0}}$ and 
$\phi^{GN}_{-}(\omega) = (1/\omega_{0})
e^{-\omega/\omega_{0}}$
with $\omega_{0} = 2\bar{\Lambda}/3$, inspired by the QCD sum rule estimates \cite{Grozin:1997pq}.
These have very different shape compared with 
Eq. (\ref{solm}), except
the behavior $\phi_{+}^{GN}(\omega) \sim \omega$,
$\phi_{-}^{GN}(\omega) \sim {\rm const}$,
as $\omega \rightarrow 0$.

Next we proceed to the case (ii), where we get
\bea
\label{eq:1}
\lefteqn{\omega \frac{\partial \Phi_{-}}{\partial \omega}+ \Phi_{+}
+ z^2\frac{\partial}{\partial z^2}
\left(\Phi_{+}
-\Phi_{-}\right) 
= 0\ ,} \\
\label{eq:2}
\lefteqn{\left(\omega 
\frac{\partial}{\partial \omega} + 2\right)\left(\Phi_{+} -
\Phi_{-}\right)
+ 4  \frac{\partial^{2}}{\partial \omega^{2}}\frac{\partial
\Phi_{+}}{\partial z^2}
= 0\ ,} \\
\label{eq:3}
\lefteqn{\left[(\omega - \bar{\Lambda}) \frac{\partial}{\partial \omega}\!+\!\frac{3}{2}\right]\Phi_{+}
\!-\!\frac{1}{2}\Phi_{-}
\! + 2  \frac{\partial^{2}}{\partial \omega^{2}}\frac{\partial \Phi_{+}}{\partial
z^2}
\!= \!0,} \\
\label{eq:4}
\lefteqn{\left[ (\omega - \bar{\Lambda})\frac{\partial}{\partial \omega}+ 2\right]
 \left(\Phi_{+}
- \Phi_{-}\right)}\nonumber\\ 
&&\;\;\;\;\;\;\;\;\;\;\;\;\;\;\;+
2  \frac{\partial^{2}}{\partial \omega^{2}}
\left(\frac{\partial \Phi_{+}}{\partial z^2} -
\frac{\partial \Phi_{-}}{\partial z^2} \right)
= 0\ ,
\eea
corresponding to Eqs. 
(\ref{de1}), 
(\ref{de2}), 
(\ref{de3}), 
(\ref{de4}),
respectively. 
Here 
Eqs. (\ref{eq:1})-(\ref{eq:4}) are given in the ``$\omega$-representation'' 
$\Phi_{\pm} \equiv \Phi_{\pm}(\omega, z^2)$,
instead of the ``$t$-representation'',
via $\tilde{\Phi}_{\pm}(t, z^{2}) = \int d\omega \ e^{-i \omega t}
\Phi_{\pm}(\omega, z^{2})$.
The light-cone limit is not taken
so that the terms proportional
to $z^{2}$  appear in Eq. (\ref{eq:1}).
Note that we 
have neglected
the contribution 
from the quark-antiquark-gluon
three-particle operators.

By combining Eqs. (\ref{eq:2}) and (\ref{eq:3}), 
we eliminate the last term in their LHS.
The resulting equation
can be integrated with boundary conditions
$\Phi_{\pm}(\omega, z^{2}) = 0$ for  $\omega <0$ or $\omega \rightarrow
\infty$ as
\begin{equation}
\left( \omega - 2 \bar{\Lambda} \right) \Phi_{+} + \omega \Phi_{-} = 0\; .
\label{eq:t2}
\end{equation}
{}In the limit $z^{2} \rightarrow 0$, Eqs. (\ref{eq:1}) and (\ref{eq:t2}) 
reduce to Eqs. (\ref{mde1}) and (\ref{mde2}) in the Wandzura-Wilczek approximation. 
The corresponding solution (\ref{solm})
serves as ``boundary conditions''
to solve Eqs. (\ref{eq:1})-(\ref{eq:4}) for $z^{2} \neq 0$.
Then, from Eqs. (\ref{eq:1}) and (\ref{eq:t2}), we find,
as the solution in the Wandzura-Wilczek approximation 
for $z^{2} \neq 0$,
\begin{equation}
\Phi^{(W)}_{\pm}(\omega, z^{2}) = \phi_{\pm}^{(W)}(\omega)
\xi \left(z^{2}\omega(2\bar{\Lambda}-\omega)\right)\ .
\label{eq:chi}
\end{equation}
Here $\xi(x)$ is a function of a single variable $x$, 
and can be 
determined from a remaining
differential equation (\ref{eq:3}) or (\ref{eq:4}) as
$\xi(x) = J_{0}\left(\sqrt{-x}\right)$
where $J_{0}$ is a (regular) Bessel function.
This result gives analytic solution for the light-cone wavefunctions with the transverse separation
$\mbox{\boldmath $z$}_{T}^2 = -z^{2}$.
The momentum-space wavefunctions 
$\Phi^{(W)}_{\pm}(\omega, \mbox{\boldmath
$k$}_{T})$,
defined by
$\tilde{\Phi}^{(W)}_{\pm}(t, -\mbox{\boldmath $z$}_{T}^2) =
\int d\omega d^{2}k_{T}\
e^{-i\omega t + i\mbox{\boldmath $k$}_{T}\cdot\mbox{\boldmath $z$}_{T}}
\Phi^{(W)}_{\pm}(\omega, \mbox{\boldmath $k$}_{T})$,
read
\be
\label{eq:12}
\Phi^{(W)}_{\pm}(\omega, \mbox{\boldmath $k$}_{T}) =
\frac{\phi_{\pm}^{(W)}(\omega)}{\pi}
\delta \left(\mbox{\boldmath $k$}_{T}^{2} - \omega (2 \bar{\Lambda} -
\omega) \right) .
\ee
The result (\ref{eq:12}) 
gives exact description
of the valence Fock components of the $B$ meson
wavefunctions in the heavy-quark limit, and represent their transverse
momentum dependence explicitly.
These results show that the dynamics within the two-particle Fock states
is determined solely in terms of a single nonperturbative
parameter $\bar{\Lambda}$.

The transverse momentum distributions in the $B$ mesons
have been 
unknown,
so that various models have been used in the literature.
Frequently used models assume
complete separation (factorization)
between the longitudinal and transverse momentum-dependence in the
wavefunctions (see e.g. Refs. \cite{kls,sachrajda}).
A typical example of such models \cite{kls} is given by
$\Phi_{\pm}^{KLS}(\omega, \mbox{\boldmath $k$}_{T})
= N\omega^{2}(1-\omega)^{2}e^{-\omega^{2}/(2\beta^{2})}
\times e^{- \mbox{\boldmath $k$}_{T}^{2}/(2\kappa^{2})}$
with some constants $N, \beta$, and $\kappa$.
Eq. (\ref{eq:12}) 
shows that 
transverse and longitudinal momenta are strongly 
correlated through the combination
$\mbox{\boldmath $k$}_{T}^{2}/[\omega(2\bar{\Lambda}-\omega)]$,
therefore the ``factorization models''
are not justified.
We further note that many models assume 
Gaussian distribution
for the $\mbox{\boldmath $k$}_{T}$-dependence as in $\Phi_{\pm}^{KLS}$.
These models 
show strong dumping at 
large $|\mbox{\boldmath $z$}_{T}|$ as $\sim \exp\left(-
\kappa^2\mbox{\boldmath $z$}_{T}^{2}/2\right)$.
In contrast,
our 
wavefunctions 
(\ref{eq:chi}) 
have slow-damping with oscillatory behavior
as
$\Phi^{(W)}_{\pm}(\omega, -\mbox{\boldmath $z$}_{T}^{2})
\sim \cos(|\mbox{\boldmath $z$}_{T}|\sqrt{\omega(2\bar{\Lambda}-\omega)}
-\pi/4)
/\sqrt{|\mbox{\boldmath $z$}_{T}|}$.

Finally, we can 
estimate the 
effects neglected in our solution (\ref{eq:12}).
Inspecting the $t \rightarrow 0$ limit 
of Eqs. (\ref{de2}), (\ref{de4}),
one immediately obtains the exact result for 
$\partial \tilde{\phi}_{\pm}(0)/\partial z^{2}$, which gives
the first moment of $\mbox{\boldmath $k$}_{T}^{2}$
as \cite{KKQT2}
\begin{equation}
\int \!\!d\omega d^{2}k_{T}\ \mbox{\boldmath $k$}_{T}^{2}
\Phi_{\pm}(\omega, \mbox{\boldmath $k$}_{T})
= \frac{2}{3}\left(\bar{\Lambda}^{2}+\lambda_{E}^{2}+\lambda_{H}^{2}\right)\! ,
\label{eq:3part}
\end{equation}
with $\lambda_{E}$ and $\lambda_{H}$ of Eq. (\ref{mome2}).
Here $\Phi_{\pm}= \Phi^{(W)}_{\pm}+\Phi_{\pm}^{(g)}$ denote
the total wavefunctions which include the higher Fock contributions $\Phi_{\pm}^{(g)}$
induced by the three-particle operators.
{}From Eq. (\ref{eq:12}), we get
$\int d\omega d^{2}k_{T}\ \mbox{\boldmath $k$}_{T}^{2}
\Phi_{\pm}^{(W)}(\omega, \mbox{\boldmath $k$}_{T})
= 2\bar{\Lambda}^{2}/3$, so that
the terms $2(\lambda_{E}^{2}+ \lambda_{H}^{2})/3$ of Eq. (\ref{eq:3part}) 
come from 
$\Phi_{\pm}^{(g)}$.
The result (\ref{eq:3part}), combined with a QCD sum rule estimate of $\lambda_{E}, \lambda_{H}$
mentioned below Eq. (\ref{mome2}),
suggests that the higher Fock 
contributions might considerably broaden the transverse momentum distribution.
However, qualitative
features discussed above, like non-factorization of
longitudinal and transverse directions, ``slow-damping'' for transverse
directions, etc.,
will be unaltered by the effects of multi-particle states.

To summarize, 
we have derived a system of differential equations
for the $B$ meson light-cone wavefunctions 
and obtained the corresponding analytic solution. 
The differential equations are derived from the exact equations 
of motion of QCD in the heavy-quark limit.
The heavy-quark symmetry plays an essential role
by reducing the number of 
independent wavefunctions drastically,  
so that the configuration of quark and antiquark in 
the $B$ mesons is described by
only two light-cone wavefunctions.
As a result, a system of four differential equations from the 
equations of motion
allows us to obtain model-independent
formulae of these two wavefunctions, which reveal roles
of the leading 
Fock-states,
as well as
the 
higher Fock-states
with additional dynamical gluons.
%
Also due to
the power of heavy-quark symmetry,
our Wandzura-Wilczek parts (\ref{solm}), (\ref{eq:12}), which correspond to the leading Fock-states,
are given in simple analytic formulae involving
one single nonperturbative parameter $\bar{\Lambda}$.
Heavy-quark symmetry also guarantees that our solutions 
determine 
the light-cone 
wavefunctions
for the $B^{*}$ mesons and also for the $D$, $D^{*}$ mesons in the
heavy-quark limit.

We emphasize that our solutions 
provide the powerful framework for building up the $B$ meson
light-cone 
wavefunctions
and their phenomenological applications,
because the solutions satisfy all relevant QCD constraints.
{}Further developments like those required 
to clarify the effects of multi-particle states 
can be exploited systematically starting from the exact results in this
work.


\end{document}